\documentclass[aps,prl,twocolumn,superscriptaddress,showpacs]{revtex4}
\usepackage{graphics}

\bibliography{20}

\begin{document}

\title{Geometrical Optics of Beams with Vortices: Berry Phase and Orbital
Angular Momentum Hall Effect}
\author{Konstantin~Yu.~Bliokh}
\affiliation{Institute of Radio Astronomy, Kharkov, 61002, Ukraine}
\affiliation{A.Ya.~Usikov Institute of Radiophysics and Electronics, Kharkov, 61085,
Ukraine}
\affiliation{Department of Physics, Bar-Ilan University, Ramat Gan, 52900, Israel}

\begin{abstract}
We consider propagation of a paraxial beam carrying the spin angular
momentum (polarization) and intrinsic orbital angular momentum (IOAM) in a
smoothly inhomogeneous isotropic medium. It is shown that the presence of
IOAM can dramatically enhance and rearrange the topological phenomena that
previously were considered solely in connection to the polarization of
transverse waves. In particular, the appearance of a new-type Berry phase
that describes the parallel transport of the beam structure along a curved
ray is predicted. We derive the ray equations demonstrating the splitting of
beams with different values of IOAM. This is the orbital angular momentum
Hall effect, which resembles Magnus effect for optical vortices. Unlike the
recently discovered spin Hall effect of photons, it can be much larger in
magnitude and is inherent to waves of any nature. Experimental means to
detect the phenomena is discussed.
\end{abstract}

\pacs{41.20.Jb, 03.65.Vf, 41.85.-p, 42.15.-i}
\maketitle

\paragraph{Introduction.}

The total angular momentum (TAM) of an electromagnetic wave packet (or a
beam) can be presented in the form of three summands: TAM=EOAM+IOAM+SAM.
Here IOAM stands for the intrinsic orbital angular momentum of the packet,
i.e. the orbital angular momentum relative to the center of gravity of the
packet, EOAM is the extrinsic orbital angular momentum related to the motion
of the center of gravity, and SAM is the spin angular momentum determined by
the polarization of the wave packet. While EOAM and SAM are well known and
studied during many decades, the beams carrying IOAM evoked intensive
theoretical and experimental investigations only in last 14 years \cite{1}.
This fact is connected to the contemporary laboratory resources for the
generation and transformation of light beams as well as single photons with
non-zero values of IOAM. The most popular beams with IOAM are the
Laguerre--Gaussian beams which contain the optical vortices along their axes
and form a complete, orthogonal, basic set from which an arbitrary field
distribution can de described. The states of light with well-defined IOAM
offer now exciting possibilities for the optical manipulation with matter,
study of the entanglement of photons, and a lot of problems of the classical
electrodynamics \cite{1}.

Simultaneously, in the past decade, the topological phenomena related to the
spin of particles caused a great activity in the various areas of physics:
condensed matter, high energy physics, optics. It is caused by both
fundamental character of the problems and promising applications in new
areas of nanotechnologies, such as spintronics, photonics, etc. Nontrivial
evolution of the spin states in the semiclassical approximation can be
treated by means of the spin-orbit interaction, which is the coupling
between SAM and EOAM. Spin-orbit interaction leads to the mutual influence
of the polarization and the trajectory of the particle's motion, and
produces two reciprocal topological phenomena. These are the geometrical
Berry phase \cite{2,3} and recently discovered topological spin transport or
intrinsic spin Hall effect \cite{4}. In the geometrical optics, the Berry
phase provides for the parallel transport of the polarization vector along
the ray \cite{3}, whereas the spin Hall effect manifests itself as the
transverse deflection of polarized beams when propagating in an
inhomogeneous medium. The latter phenomenon includes the
polarization-dependent transverse Fedorov--Imbert shift in the reflection or
refraction of the beam on a sharp boundary \cite{5,6,8,10}, and the
splitting of rays of different polarizations in a smoothly inhomogeneous
medium \cite{6,7,8,9}. The common roots of these effects and their
connection to the spin-orbit interaction and TAM conservation are indicated
in the papers \cite{6,8,10}.

All mentioned phenomena are related to the interaction between the
polarization of light (SAM of the field) and its extrinsic orbital features,
EOAM. At the same time, the presence of the IOAM in the beam apparently may
lead to analogous effects. Then large values of IOAM can give a great
advantage compared to small polarization phenomena constrained by unit value
of SAM per photon ($\hbar=1$). The transverse Fedorov--Imbert shift related
to IOAM has been recently described and measured \cite{11}. In the present
paper, similarly to the developed earlier geometrical optics with the
spin-orbit interaction \cite{6,7,8,9}, we evolve the geometrical optics with
the \textit{orbit-orbit interaction}, i.e., an interaction between IOAM and
EOAM of the beam propagating along a curved ray.

\paragraph{Berry connection and curvature.}

We consider a monochromatic paraxial electromagnetic beam with definite
values of SAM and IOAM, which propagates in a smoothly inhomogeneous
isotropic medium. Locally, the beam's electric field (without $e^{-i\omega
t} $ factor) can be represented as in homogeneous medium: 
\begin{equation}  \label{eq1}
\mathbf{E}^{p,l,\sigma } = \mathbf{e}^\sigma F^{p,\left| l \right|} \left(
\rho \right)\exp \left[ {il\varphi + i\int {k} d{s}} \right].
\end{equation}
Here $\mathbf{k}(s)$ is the central wave vector directed along the beam $s$ axis; $%
\left(\rho,\varphi,s\right)$ are the local cylindrical coordinates following
the central ray whose trajectory is described by the geometrical optics ray
equations; $\mathbf{e}^\sigma$ is the unit vector of the polarization of
wave with the helicity $\sigma=\pm 1$ which is the value of $s$-directed SAM
per photon ($\mathbf{e}^{\pm}$ correspond to waves of right and left
circular polarizations, so that $\mathbf{e}^{\pm\ast}=\mathbf{e}^{\mp}$ and $%
\mathbf{e}^{\sigma\dag} \mathbf{e}^{\sigma ^{\prime}} = \delta ^{\sigma
\sigma ^{\prime}}$); $F^{p,\left| l \right|}$ is the radial function with
quantum number $p = 0,1,2,...$; finally, $l = 0, \pm 1,..., \pm p$ is the
azimuthal quantum number which is the value of $s$-directed IOAM per photon.
The expression (\ref{eq1}) is given in a diffractionless approximation and
does not account for variations of the beam's envelope, phase front, and the
Guoy phase; as it is known, the diffraction phenomena do not contribute to
the geometrical optics characteristics of the wave packet evolution. In the
paraxial approximation one can consider the longintudinal wave vector $k$ as
independent of the transverse structure of the wave field and the
polarization vector $\mathbf{e}^\sigma$ as common for the whole beam. This
is the paraxial approximation that enables one to consider the states of
field with simultaneously well-defined values of the helicity (SAM) and IOAM %
\cite{1}. The axis of the beam (\ref{eq1}) contains an optical dislocation
(the phase singularity): the optical vortex of the strength $l$. The scalar
product for beams (\ref{eq1}) is defined as for vectors in Hilbert space,
trough integration over the beam cross-section with $s$ replacing `time' %
\cite{1}; we assume that beams (\ref{eq1}) form an orthonormal basis: $%
\left( {\mathbf{E}^{p,l,\sigma } ,\mathbf{E}^{p^{\prime},l^{\prime},\sigma
^{\prime}} } \right) \equiv \int\int {\mathbf{E}^{p,l,\sigma\dag} \mathbf{E}%
^{p^{\prime},l^{\prime},\sigma ^{\prime}}\rho d {\rho}d {\varphi}} = \delta
^{pp^{\prime}} \delta ^{ll^{\prime}} \delta ^{\sigma \sigma ^{\prime}}$.

The variations in the ray direction, $\mathbf{k}$, give rise to a nontrivial
parallel transport law. It is determined by the Berry connection (gauge
potential) defined as 
\begin{equation}  \label{eq2}
\mathcal{A}^{\sigma \sigma ^{\prime}ll^{\prime}}(\mathbf{k}) = i\left( {%
\mathbf{E}^{p,l,\sigma } ,\frac{\partial }{{\partial \mathbf{k}}}\mathbf{E}%
^{p,l^{\prime},\sigma ^{\prime}} } \right)
\end{equation}
(the radial index contributes to the Berry connection in a trivial way and
is not considered). Equation (\ref{eq2}) with the field (\ref{eq1}) yields $%
\mathcal{A}^{\sigma \sigma ^{\prime}ll^{\prime}} = \delta^{ll^{\prime}}%
\mathcal{A}^{\sigma \sigma ^{\prime}} + \delta^{\sigma \sigma ^{\prime}}%
\mathcal{A}^{ll^{\prime}}$. Here $\mathcal{A}^{\sigma \sigma ^{\prime}} = i%
\mathbf{e}^{\sigma \dag} \frac{\partial }{{\partial \mathbf{k}}}\mathbf{e}%
^{\sigma ^{\prime}}$ is the known spin-related Berry connection for plane
waves that accounts for the parallel transport of the polarization vector in
the $\mathbf{k}$-space \cite{3}, whereas $\mathcal{A}^{ll^{\prime}} =
i\int\limits_0^{2\pi } {e^{ - il\varphi } } \frac{\partial }{{\partial 
\mathbf{k}}}e^{il^{\prime}\varphi } d\varphi $ is a new term connected to
the IOAM. The spin term is a diagonal (Abelian) gauge potential: $\mathcal{A}%
^{\sigma \sigma ^{\prime}} = \delta ^{\sigma \sigma ^{\prime}} \mathcal{A}%
^\sigma = \delta ^{\sigma \sigma ^{\prime}} \sigma \mathcal{A}$, which gives
rise to the gauge field (Berry curvature) of the magnetic monopole type: $%
\mathcal{F}^{\sigma \sigma ^{\prime}} = \delta ^{\sigma \sigma ^{\prime}} 
\mathcal{F}^\sigma = \delta ^{\sigma \sigma ^{\prime}} \sigma \mathcal{F}$, $%
\mathcal{F} = \frac{\partial }{{\partial \mathbf{k}}} \times \mathcal{A} = - 
\frac{\mathbf{k}}{{k^3 }}$ \cite{3,7,8}. By evaluating the orbit term, we
derive $\mathcal{A}^{ll^{\prime}} = \delta ^{ll^{\prime}} l\mathcal{A}$.
Finally, the total Berry connection and curvature $\mathcal{F}^{\sigma
\sigma ^{\prime}ll^{\prime}} = \frac{\partial }{{\partial \mathbf{k}}}
\times \mathcal{A}^{\sigma \sigma ^{\prime}ll^{\prime}}$ read: 
\begin{equation}  \label{eq3}
\mathcal{A}^{\sigma \sigma ^{\prime}ll^{\prime}} = \delta ^{\sigma \sigma
^{\prime}}\delta ^{ll^{\prime}}\left( {\ \sigma + l} \right)\mathcal{A},~%
\mathcal{F}^{\sigma \sigma ^{\prime}ll^{\prime}} = \delta ^{\sigma \sigma
^{\prime}}\delta ^{ll^{\prime}}\left( {\ \sigma + l} \right)\mathcal{F}.
\end{equation}

Thus, the Berry connection for beams (\ref{eq1}) is also a diagonal Abelian
potential represented by a tensor of higher rank. Propagation in a smoothly
inhomogeneous medium keeps the beam's helicity, $\sigma$, and IOAM, $l$,
conserved in the geometrical optics approximation. Hence, the Berry
connection and curvature for a given beam (\ref{eq1}) equal $\mathcal{A}%
^{\sigma l} = \left( {\sigma + l} \right)\mathcal{A}$ and $\mathcal{F}%
^{\sigma l} = \left( {\sigma + l} \right)\mathcal{F} = - \left( {\sigma + l}
\right)\mathbf{k}/k^3$. Thus, topologically, a beam with IOAM behaves
similarly to a polarized beam containing only SAM; however, the magnitude of
the phenomena changes: instead of SAM, $\sigma$, now they are proportional
to the total intrinsic angular momentum of the beam: SAM+IOAM, $\sigma+l$.
In contrast to situations considered previously, the nontrivial topological
features can be manifested by beams without SAM, $\sigma=0$, in particular
by beams of longitudinal (sound) waves with $l\neq 0$.

\paragraph{Geometrical phase and parallel transport of the beam structure.}

We proceed to describe the basic phenomena arising from the Berry connection
(\ref{eq3}). On the evolution, the beam acquires an
additional phase, namely, the Berry geometrical phase that equals 
\begin{equation}
\Theta _{B}^{\sigma l}=\int\limits_{C}{\mathcal{A}^{\sigma l}}d\mathbf{k}%
=\left( {\sigma +l}\right) \int\limits_{C}{\mathcal{A}\,}d\mathbf{k}=\left( {%
\sigma +l}\right) \Theta _{B0},  \label{eq4}
\end{equation}%
where $C$ is the contour of the ray evolution in the $\mathbf{k}$-space, and 
$\Theta _{B0}$ is the Berry phase accumulated in the beam with $\sigma =1$
and $l=0$. For a cyclic evolution in the $\mathbf{k}$-space, when $C$ is a
loop, the Berry phase can be represented as a surface integral of the Berry
curvature: $\Theta _{B}^{\sigma l}=\left( {\sigma +l}\right) \oint\limits_{S}%
\mathcal{F}d\mathbf{k}=-\left( {\sigma +l}\right) \Omega $, where $%
C=\partial S$, and $\Omega $ is the solid angle at which contour $C$ is seen
from the origin $\mathbf{k}=0$. With the Berry phase (\ref{eq4}) taken into
account, the field (\ref{eq1}) of the beam propagating along a curved ray
should be rewritten as 
\begin{equation}
\mathbf{E}^{p,l,\sigma }=\mathbf{e}^{\sigma }e^{i\sigma \Theta
_{B0}}F^{p,\left| l\right| }\exp \left[ {il\left( {\varphi +\Theta _{B0}}%
\right) +i\int {k} d{s}}\right] .  \label{eq5}
\end{equation}%
This equation is one of the central results of the paper. There factor $%
e^{i\sigma \Theta _{B0}}$ is the known Berry phase associated with the
parallel transport of the polarization vector \cite{3}, while the new-type
`orbital' Berry phase, incorporated into the azimuthal distribution, shows
that \textit{on the evolution, the transverse distribution of the field
rotates at the angle $-\Theta _{B0}$, which corresponds to the parallel
transport of the beam structure along the ray}. It follows from Eq.~(\ref%
{eq5}) that an arbitrary electromagnetic beam representable as a
superposition of beams (\ref{eq1}), e.g. a Hermite--Gaussian beam, will also
experience a rotation at the angle $-\Theta _{B0}$ in accordance with the
parallel transport law, Fig. 1 \cite{12}. The rotation of the transverse
intensity distribution in the beam can be observed in propagation of a
Hermite--Gaussian beam along a helical ray in an axially symmetric medium or
in a circular multimode optical fiber stranded in a helix. In the latter
case, the eigen modes of the fiber have the form of Eq. (\ref{eq1}) \cite%
{13,14} and, hence, the speckle pattern at the output of the fiber will
rotate depending on its torsion \cite{15}. In the fiber, the eigen modes (%
\ref{eq1}) experience no diffraction, which will be helpful when detecting
the effect \cite{12}.

\begin{figure}[t]
\centering \scalebox{0.55}{\includegraphics{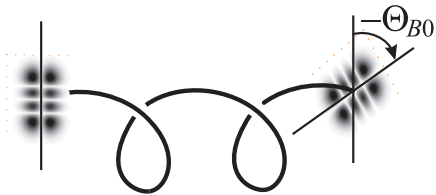}}
\caption{The parallel transport of the cross-structure of a beam ($\text{HG}%
_{31}$-mode here) propagating along a helical trajectory.}
\label{Fig1}
\end{figure}

\paragraph{Ray equations: spin and orbital Hall effects.}

The ray equations, which describe the motion of the center of the beam (\ref%
{eq5}) in a smoothly inhomogeneous medium, are Hamiltonian semiclassical
equations of motion. They have been derived a number of times for various
particles in the presence of the Berry curvature in the $\mathbf{k}$-space %
\cite{2,4,6,7,8,9}. Completely similarly to the case of the spin-related
Berry curvature \cite{6,7,8,9}, we obtain: 
\begin{equation}
\mathbf{\dot{k}}=k\nabla \ln n,~\mathbf{\dot{r}}=\frac{\mathbf{k}}{k}+%
\mathcal{F}^{\sigma l}\times \mathbf{\dot{k}}=\frac{\mathbf{k}}{k}-\left( {%
\sigma +l}\right) \left( {\frac{\mathbf{k}}{{k^{3}}}\times \mathbf{\dot{k}}}%
\right) ,  \label{eq6}
\end{equation}%
where dot denotes the derivative with respect to $s$, and $n(\mathbf{r})$ is
the refractive index of the medium. Equations (\ref{eq6}) are the second
central results of the paper. They differ from the `traditional' ray
equations of the geometrical optic by the term proportional to the Berry
curvature in the right-hand side of the second equation. This term is
referred to as ``anomalous velocity'' (since it contributes to $\mathbf{\dot{%
r}}$) \cite{4}; it gives rise to the ray deflection, $\delta \mathbf{r}%
^{\sigma l}$, which crucially depends on the polarization and IOAM of the
beam, Fig. 2. When the anomalous velocity does not contribute to $\mathbf{k}$%
, the deflection, like the Berry phase, can be represented in the form of a
contour integral in the $\mathbf{k}$-space: $\delta \mathbf{r}^{\sigma
l}=\int\limits_{C}{\mathcal{F}^{\sigma l}\times }d\mathbf{k}=-\left( {\sigma
+l}\right) \int\limits_{C}{\frac{{\mathbf{k}\times d\mathbf{k}}}{{k^{3}}}}$.
Here contour $C$ is determined by the rays of zero approximation, i.e. by
Eqs. (\ref{eq6}) with $\mathbf{\dot{r}}=\mathbf{k}/k$. The deflection is
small in magnitude (it is proportional to the wavelength) but can grow
unlimitedly with the ray length $s$; its maximum value can be estimated as $%
\left| {\delta \mathbf{r}^{\sigma l}}\right| _{\max }\sim \left| {\sigma +l}%
\right| s/kL$, where $L$ is the characteristic scale of the medium
inhomogeneity. Since the anomalous velocity is directed orthogonally to both
the main beam propagation direction, $\mathbf{k}$, and the applied `external
force' $\mathbf{\dot{k}}\propto \nabla n$ (Fig. 2), the phenomenon can be
treated as intrinsic Hall effect of photons, which is of a universal nature
for various particles \cite{4}. If one separates the spin and orbital parts
in an obvious way, $\delta \mathbf{r}^{\sigma l}=\delta \mathbf{r}^{\sigma
}+\delta \mathbf{r}^{l}$, the previously known SAM-related deflection $%
\delta \mathbf{r}^{\sigma }$ can be associated with the spin Hall effect of
photons \cite{6,7,8,9}, whereas the IOAM-related deflection $\delta \mathbf{r%
}^{l}$, introduced in the present paper, represents a novel effect -- 
\textit{intrinsic orbital angular momentum Hall effect}. Indeed, the
splitting of beams (\ref{eq5}) with different (opposite) values of $l$
implies the appearance of the transverse current of IOAM (Fig.~2) \cite{16}.
Various examples of media, where the spin Hall effect gives rise to
noticeable transport of rays, can be found in papers \cite{6,7,8}; due to
mathematical similarity of ray equations, the IOAM Hall effect in those
media can be considered analogously. We emphasize that the presence of IOAM
dramatically enhances the topological transport described by Eqs.~(\ref{eq6}%
), since the value of $l$ can reach tens and even hundreds. The IOAM Hall
effect can be much more efficient as compared with the spin Hall effect and
can be observed independently on the wave polarization.

Since the center of the beam (\ref{eq5}) is a vortex core, Eqs. (\ref{eq6})
can be regarded as the equations for the field dislocation line in an
inhomogeneous medium. Then, deflection $\delta \mathbf{r}^{\sigma l}$
strongly resembles the effect of Magnus force acting on a vortex in
superconductors, where it also directly related to the Berry phase but in $%
\mathbf{r}$- rather than $\mathbf{k}$-space \cite{17} (see also \cite{14}).
Association with the vortex line gives a good opportunity for measuring $%
\delta \mathbf{r}^{\sigma l}$, since the singularity can be observed with a
great accuracy.

\paragraph{TAM conservation and Fedorov--Imbert shift.}

TAM of the paraxial beam (\ref{eq1}) (including EOAM) can be represented in
the form $\mathbf{J} = \mathbf{r} \times \mathbf{k} + \left( {\sigma + l}
\right)\mathbf{k}/k$. It can be easily shown (see \cite{4,8}), that it is the
Berry curvature term in the equations of motion (\ref{eq6}) that guarantees
conservation of $J_z$ in an axially symmetrical with respect to $z$ axis
medium, as well as conservation of $\mathbf{J}$ in a medium spherically
symmetric with respect to the origin. The conservation of TAM of the beam
reveals the common nature of the transport under consideration and the
transverse Fedorov--Imbert shift at a sharp interface between two media. The
latter problem has been analyzed rigorously for paraxial beams with IOAM in %
\cite{11}. When a Laguerre--Gaussian beam experiences scattering at the
interface with a low contrast $\delta n \ll 1$, the reflected beam can be
neglected, and the transverse shift of the refracted beam approximately
equals \cite{11} $\delta x \approx \frac{{\left( {\sigma + l} \right)}}{k}%
\frac{{\delta n}}{n}\tan \vartheta $. Here $\vartheta$ is the angle of
incidence, and $x$ axis points orthogonally to both $\mathbf{k}$ and the
normal to the interface ($\nabla n$). Precisely the above shift provides for
conservation of the normal component of $\mathbf{J}$ \cite{5,8,10}. In the
differential limit, $\delta n \to dn$, $\delta x \to dx$, it leads to the
ray equations (\ref{eq6}) in a smoothly inhomogeneous medium, see \cite{6}.

\paragraph{Evolution of beams superposition.}

Suppose now that the field presents a superposition of beams (\ref{eq1})
with different quantum numbers: $\mathbf{E}=A\sum\limits_{p,l,\sigma }{%
a^{pl\sigma }\mathbf{E}^{p,l,\sigma }}$, where the common amplitude $A$ is
chosen in such a way that $\sum\limits_{p,l,\sigma }\left| a^{pl\sigma
}\right| ^{2}=1$. A set of amplitudes $a^{pl\sigma }$ represents the unit
state vector in the basis of beams (\ref{eq1}) \cite{1}. Then, one can
determine the expected value of any quantity for the superposition state.
Since the Berry gauge field is Abelian, the transition to the expected
values is realized in a straightforward way. The ray equations for expected
values, $\mathbf{\bar{r}}$ and $\mathbf{\bar{k}}$, (which are the
coordinates of the center of gravity of the total beam in the phase space),
takes the form similar to Eqs.~(\ref{eq6}) where one component of the Berry
curvature $\mathcal{F}^{\sigma l}$ is replaced with the convolution $\bar{%
\mathcal{F}}=\sum\limits_{p,l,l^{\prime },\sigma ,\sigma ^{\prime }}{%
a^{pl\sigma \ast }\mathcal{F}^{\sigma \sigma ^{\prime }ll^{\prime
}}a^{pl^{\prime }\sigma ^{\prime }}}=\sum\limits_{p,l,\sigma }{\left| {%
a^{pl\sigma }}\right| ^{2}\mathcal{F}^{\sigma l}}$, and quantum numbers $%
\sigma $ and $l$ are replaced by `classical' quantities $\bar{\sigma}%
=\sum\limits_{p,l,\sigma }{\left| {a^{pl\sigma }}\right| ^{2}\sigma }$ and $%
\bar{l}=\sum\limits_{p,l,\sigma }{\left| {a^{pl\sigma }}\right| ^{2}l}$. It
is worth noticing that though the evolution of the beam's center of gravity
is described by the expected values, in actual fact the total beam splits
into the partial beams propagating along slightly shifted trajectories (\ref%
{eq6}). This can be detected, as in the Hall effects, through accumulation
of the photon states with opposite signs of $\sigma $ or $l$ at the opposite
sides of the beam, see \cite{10,18}, or through the splitting of the
singularity lines of vortices with different $l$. In this way, a
Hermite--Gaussian beam propagating in an inhomogeneous medium will split
into composing Laguerre--Gaussian beams (\ref{eq1}) with opposite values of $%
l$ (the birth of vortex pairs will accompany this process). Like the
circularly rather than linearly polarized plane waves are independent modes
in an inhomogeneous medium \cite{7,10,18}, it is the Laguerre--Gaussian
beams that constitute the independent localized modes. This follows from the
diagonality of the Berry connection (\ref{eq3}) in the basis of beams (\ref%
{eq1}).

When a superposition of beams (\ref{eq1}) is considered, their Berry phases
introduced explicitly in Eq. (\ref{eq5}) can be taken into account by means
of the equation of evolution of the state vector $a^{pl\sigma }$.
Analogously to the equation of motion for the state vector with one quantum
number $\sigma$ \cite{8}, in general case we have: 
\begin{equation}  \label{eq7}
\dot a^{pl\sigma } = i\sum\limits_{l^{\prime},\sigma ^{\prime}} {\mathcal{A}%
^{ll^{\prime}\sigma \sigma ^{\prime}} \mathbf{\dot k}a^{pl^{\prime}\sigma
^{\prime}} } = i\mathcal{A}^{l\sigma } \mathbf{\dot k}a^{pl\sigma }.
\end{equation}
Integration of Eq.~(\ref{eq7}) gives the Berry phases acquired by the
partial beams, $a^{pl\sigma } = a_0^{pl\sigma } \exp \left[ {i\left( {\sigma
+ l} \right)\Theta _{B0}} \right]$, and the parallel transport of the beam's
field.

\begin{figure}[t]
\centering \scalebox{0.55}{\includegraphics{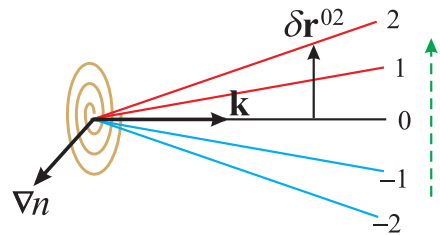}}
\caption{(color online). Transverse transport of rays (vortex cores) in an
inhomogeneous medium. The rays are marked with values of $\protect\sigma+l$.
The dashed arrow points the direction of the intrinsic angular momentum
current.}
\label{Fig2}
\end{figure}

\paragraph{Conclusion.}

We have considered propagation of a paraxial beam carrying intrinsic spin
and orbital angular momenta in a smoothly inhomogeneous medium in the
geometrical optics approximation. It is shown that expressions for the Berry
connection and curvature, as well as the equations of motion, are similar to
situation when only SAM is presented, with a substitution $\sigma \to \sigma
+ l$. However, there are some sharp distinctions. First, the additional
Berry phase acquired during the beam evolution provides for the parallel
transport of the intensity cross distribution in the beam along the ray. The
effect can be detected when a Hermite--Gaussian beam propagates of along a
helical trajectory or through the measurement of rotation of the
speckle-pattern when a circular multi-mode fiber is twisted. Second, the ray
equations describing deflections of beams with different values of $\sigma +
l$, predicts thereby the intrinsic orbital angular momentum Hall effect for
photons, which resembles Magnus effect for optical vortices. The effect has
significant advantages compared to the spin Hall effect of photons: (i) it
can be dozen of times larger in magnitude with the respective values of $l$;
(ii) it is independent of the polarization and takes place for waves of any
nature. The effect can be measured through the deflection or splitting of
singularity lines for vortices with different $l$.


\begin{references}
\bibitem{1}{{\it Optical Angular Momentum}, edited by L. Allen, S.M. Barnett, and M.J. Padgett (Taylor and Francis, 2003).}
\bibitem{2}{M.V. Berry, Proc. R. Soc. Lond. A {\bf 392}, 45 (1984); A. Bohm et al., {\it Geometrical Phase in Quantum Systems} (Springer Verlag, 2003).}
\bibitem{3}{S.I. Vinnitski et al., Usp. Fiz. Nauk {\bf 160}(6), 1 (1990) [Sov. Phys. Usp. {\bf 33}, 403 (1990)].}
\bibitem{4}{S. Murakami, N. Nagaosa, and S.-C. Zhang, Science {\bf 301}, 1348 (2003); D. Culcer et al., Phys. Rev. Lett. {\bf 93}, 046602 (2004); F. Zhou, Phys. Rev. B {\bf 70}, 125321 (2004); A. B\'{e}rard and H. Mohrbach, Phys. Lett. A {\bf 352}, 190 (2006); K.Yu. Bliokh and Yu.P. Bliokh, Ann. Phys. (N.Y.) {\bf 319}, 13 (2005).}
\bibitem{5}{F.I. Fedorov, Dokl. Akad. Nauk SSSR {\bf 105}, 465 (1955); C. Imbert, Phys. Rev. D {\bf 5}, 787 (1972); M.A. Player, J. Phys. A: Math. Gen. {\bf 20}, 3667 (1987); V.G. Fedoseyev, \textit{ibid.} {\bf 21}, 2045 (1988); W. Nasalski, J. Opt. Soc. Am. A {\bf 13}, 172 (1996).}
\bibitem{6}{V.S. Liberman and B.Ya. Zel'dovich, Phys. Rev. A {\bf 46}, 5199 (1992).}
\bibitem{7}{K.Yu. Bliokh and Yu.P. Bliokh Phys. Rev. E {\bf 70}, 026605 (2004); Phys. Lett. A {\bf 333}, 181 (2004); K.Yu. Bliokh and V.D. Freilikher, Phys. Rev. B {\bf 72}, 035108 (2005).}
\bibitem{8}{M. Onoda, S. Murakami, and N. Nagaosa, Phys. Rev. Lett. {\bf 93}, 083901 (2004).}
\bibitem{9}{C. Duval, Z. Horv\'{a}th, and P.A. Horv\'{a}thy, math-ph/0509031; cond-mat/0509636 (to appear in Phys. Rev D).}
\bibitem{10}{K.Yu. Bliokh and Yu.P. Bliokh, Phys. Rev. Lett. {\bf 96}, 073903 (2006).}
\bibitem{11}{V.G. Fedoseyev, Opt. Comm. {\bf 193}, 9 (2001); R. Dasgupta and P.K. Gupta, \textit{ibid.} {\bf 257}, 91 (2006).}
\bibitem{12}{The field structure obeys the parallel transport law in diffractionless approximation. Diffraction deformations of the beam envelope, in general case, do not follow the parallel transport, P. Berczynski et al., J. Opt. Soc. Am. A {\bf 23}, 1442 (2006). Screw form of the astigmatic beams, J.A. Arnaud and H. Kogelnik, Appl. Opt. {\bf 8}, 1687 (1969), can also be attributed to the diffraction phenomena.}
\bibitem{13}{A.W. Snyder and J.D. Love, {\it Optical Waveguide Theory} (Methuen, London, 1984).}
\bibitem{14}{A.V. Dooghin et al., Phys. Rev. A {\bf 45}, 8204 (1992).}
\bibitem{15}{Unlike the small (proportional to the wavelength) effect of the rotation of the speckle-pattern in a direct fiber depending on the polarization \cite{14,6}, (which can be attributed to weak spin-orbit interaction between SAM and IOAM), the proposed effect is of the order of unity and depends only on the torsion of fiber.}
\bibitem{16}{B.A. Bernevig, T.L. Hughes, and S.-C. Zhang, Phys. Rev. Lett. {\bf 95}, 066601 (2005).}
\bibitem{17}{P. Ao and D.J. Thouless, \textit{ibid.} {\bf 70}, 2158 (1993).}
\bibitem{18}{N.N.~Punko and V.V.~Filippov, Pis'ma v ZhETF {\bf 39}, 18 (1984) [JETP Letters 39, 20 (1984)].}

\end{references}
\end{document}